# Observations on Factors Affecting Performance of MapReduce based Apriori on Hadoop Cluster

Sudhakar Singh
Department of Computer Science
Institute of Science, BHU
Varanasi, India
sudhakarcsbhu@gmail.com

Rakhi Garg
Department of Computer Science
Mahila Mahavidyalaya, BHU
Varanasi, India
rgarg@bhu.ac.in

P. K. Mishra
Department of Computer Science
Institute of Science, BHU
Varanasi, India
mishra@bhu.ac.in

*Abstract*—Designing fast and scalable algorithm for mining frequent itemsets is always being a most eminent and promising problem of data mining. Apriori is one of the most broadly used and popular algorithm of frequent itemset mining. Designing efficient algorithms on MapReduce framework to process and analyze big datasets is contemporary research nowadays. In this paper, we have focused on the performance of MapReduce based Apriori on homogeneous as well as on heterogeneous Hadoop cluster. We have investigated a number of factors that significantly affects the execution time of MapReduce based Apriori running on homogeneous and heterogeneous Hadoop Cluster. Factors are specific to both algorithmic and non-algorithmic improvements. Considered factors specific to algorithmic improvements are filtered transactions and data structures. Experimental results show that how an appropriate data structure and filtered transactions technique drastically reduce the execution time. The non-algorithmic factors include speculative execution, nodes with poor performance, data locality & distribution of data blocks, and parallelism control with input split size. We have applied strategies against these factors and fine tuned the relevant parameters in our particular application. Experimental results show that if cluster specific parameters are taken care of then there is a significant reduction in execution time. Also we have discussed the issues regarding MapReduce implementation of Apriori which may significantly influence the performance.

*Keywords—Frequent Itemset Mining, Apriori, Heterogeneous Hadoop Cluster; MapReduce; Big Data*

I. INTRODUCTION

Frequent itemset mining on big data sets, is one of the most contemporary research in Data Mining [1] and Big Data [2]. In order to mine intelligence from big data sets, data mining algorithms are being re-designed on MapReduce framework to be executed on Hadoop cluster. Hadoop [3] is an extremely large scale and fault tolerant parallel and distributed system for managing and processing of big data. MapReduce is its computational framework. Big data is dumb if we don't have algorithm to make use of it [4]. It's an algorithm that transforms the data into valuable and precise information. Frequent itemset mining is one of the most important technique of data mining. Apriori [5] is the most famous, simple and well-known algorithm for mining frequent itemsets using candidate itemsets generation. Many parallel and distributed version of Apriori algorithm have been designed to enhance the speed and to mine large scale datasets [6-7]. These algorithms are efficient in analyzing data but not in managing large scale data. Hadoop is an excellent infrastructure that provides an integrated service of managing and processing excessive volumes of datasets. The core constituents of Hadoop are Hadoop Distributed File System (HDFS) and MapReduce [8-9]. HDFS provides scalable and fast access to its unlimited storage of data. MapReduce is a parallel programming model that provides an efficient and scalable processing of large volumes of data stored in HDFS.

An application executes as a MapReduce job on Hadoop cluster. MapReduce provides high scalability, as a job is partitioned into a number of smaller tasks to run in parallel on multiple nodes in cluster. MapReduce programming model is so simplified that programmers only need to focus on processing data rather than on parallelism related details e.g. data & task partition, load balancing etc. The performance of a MapReduce job running on Hadoop cluster can be optimized in two ways. The first one is the algorithm specific where algorithmic optimization can be incorporated directly. The second one is the cluster specific where one can adjust some parameters of cluster configurations and input size for datasets. Many techniques have been proposed to optimize the performance of Apriori algorithm on MapReduce framework. Hadoop is designed on the implicit assumption that nodes in the cluster are homogeneous. But in practice it's not always possible to have homogeneous nodes. Most of the laboratories and institutions are used to have heterogeneous machines. So it becomes essential to adopt proper strategies when running MapReduce job on heterogeneous Hadoop cluster. The performance of a MapReduce job running on Hadoop cluster is greatly affected by tuning of various parameters specific to cluster configuration. For Apriori like CPU intensive algorithms, the granularity of input split may lead to a major difference in execution times. In this paper we have incorporated two algorithm specific techniques data structures and filtered transactions in Apriori algorithm, which greatly reduce the execution time on both homogeneous and heterogeneous cluster. We have investigated some factors specific to cluster configuration to make the execution faster. Factors central to our discussion are speculative execution, performance of physical node versus virtual node, distribution of data blocks, and parallelism control using input split size.

Moreover we have also discussed the issue regarding MapReduce implementation of Apriori that quite possibly influence the execution time. We have executed the different variation of MapReduce based Apriori on our local heterogeneous Hadoop cluster and found that we can achieve faster execution by tuning the cluster specific parameters without making algorithmic improvements.

The rest of the paper is organized as follows. Section 2 introduces some fundamental concepts regarding Apriori algorithm, Hadoop cluster and MapReduce programming paradigm. Section 3 summarizes works related to optimization of Apriori on MapReduce framework and performance improvement of MapReduce job on heterogeneous clusters. Experimental platform is described in section 4. Factors affecting the performance of MapReduce based Apriori and strategies adopted to improve the performance along with the experimental results are discussed in section 5. Finally section 6 concludes the paper.

## II. BASIC CONCEPTS

### A. Apriori Algorithm

Apriori is an iterative algorithm proposed by R. Agrawal and R. Srikant [5], which finds frequent itemsets by generating candidate itemsets. Apriori name of the algorithm is based on the apriori property which states that all the subset (k-1)-itemsets of a frequent k-itemset must also be frequent [5].

Apriori first scans the database and count the support of each item, and then checks against minimum support threshold to generate set of frequent 1-itemset $L_1$. In $k^{th}$ iteration $(k \geq 2)$, candidate k-itemsets $C_k$ are generated from frequent (k-1)-itemsets $L_{k-1}$. Again entire database is scanned to count the support of candidates $C_k$ and tested against minimum support to generate frequent k-itemsets $L_k$. In each iteration, generate and test steps are being carried out until there is no possibility to generate more new candidates. Candidates $C_k$ are generated from frequent itemsets $L_{k-1}$ using joining and pruning actions. Frequent itemsets $L_{k-1}$ are conditionally joined with itself such that two itemsets of $L_{k-1}$ are joined if and only if their first $(k-2)$ items are equal and $(k-1)^{th}$ item of first itemset is lexicographically smaller than respective item of the other itemset. Pruning based on Apriori property reduces the size of candidates $C_k$ by removing infrequent itemsets.

### B. Hadoop and MapReduce

Hadoop is designed on the fundamental principle of distributing computing power to where the data is rather than movement of data as in traditional parallel and distributed computing system using MPI (Message Passing Interface). Hadoop is an extremely scalable and highly fault tolerant distributed infrastructure that automatically handles parallelization, load balancing and data distribution [10]. The core components of Hadoop are HDFS and MapReduce, which are inspired by Google's File System (GFS) [11] and Google's MapReduce model [12]. HDFS is capable of storing excessive volumes of data and fast accessing to the stored data. It provides high fault tolerance and high availability of data. Files are stored in HDFS after breaking into smaller data blocks (default block size is 64 MB). Blocks are replicated across multiple nodes in the cluster (default replication factor is 3). A Hadoop cluster works on master-slave architecture in which one node is a master node and remaining nodes are slave nodes. Master node known as NameNode controls the slave nodes known as DataNodes. Slave nodes hold all the data blocks and perform map and reduce tasks [10].

A computational application runs as a MapReduce job on input datasets residing in HDFS of Hadoop cluster. A MapReduce job consists of map and reduce tasks and both work on data in the form of *(key, value)* pairs. Map and reduce tasks are being executed by Mapper and Reducer class respectively of MapReduce framework. An additional combiner class may also be used executing reduce task and known as mini reducer. There are a number of instances of Mapper and Reducer running in parallel but a Reducer starts only when all the Mapper has been completed. Mapper takes input as assigned datasets, process it and produces a number of *(key, value)* pairs as output. These *(key, value)* pairs are assigned to Reducers after sorting and shuffling by MapReduce's underlying system. Shuffling procedure assigned key and list of values associated with this key to a particular Reducer. Reducer takes input as *(key, list of values)* pairs and produce new *(key, value)* pairs. Combiner works on the output of Mappers of one node to reduce the data transfer load from Mappers to Reducers. In MapReduce framework only a single time communication occurs when output of Mappers are being transferred to Reducers [10].

Apriori is an iterative algorithm which generates frequent k-itemsets in $k^{th}$ iteration. Corresponding to an iteration of Apriori, one has to trigger a new MapReduce job/phase. In each MapReduce phase input data is read from HDFS and frequent itemsets of previous phase from DistributedCache [9] to generate frequent itemsets of current phase.

## III. RELATED WORKS

Apriori is re-designed on MapReduce framework by many authors [13-17] but most of them are straight forward implementations. FPC (Fixed Passes Combined-counting) and DPC (Dynamic Passes Combined-counting) algorithms combine the multiple consecutive passes of SPC (Single Pass Counting) to enhance the performance [18]. SPC is a straight forward implementation of Apriori on MapReduce framework. Algorithm proposed by F. Kovacs and J. Illes [19] invokes candidate generation inside Reducer as it is used to be inside Mapper. Authors also proposed a triangular matrix data structure for separate counting of 1 and 2-itemsets in a single iteration. L. Li and M. Zhang [20] proposed a dataset distribution strategy for heterogeneous Hadoop cluster and used it on a single MapReduce phase implementation of Apriori. Honglie Yu *et al.* [21] proposed an algorithm on Hadoop that uses Boolean Matrix and *AND* operator on this matrix. A parallel randomized algorithm, PARMA proposed by Matteo Riondato *et al.* [22] discovers approximate frequent itemsets from a sample of datasets.

Many works have been done for improving MapReduce performance in heterogeneous Hadoop clusters. J. Xie *et al.* [23] developed a data placement management mechanism and incorporated two algorithms (named as Initial Data Placement and Data Distribution) into HDFS. The first algorithm divides a large input file into even-sized fragments and then assigns fragments to heterogeneous nodes in a cluster according to data processing speed of nodes. Processing speed of nodes is quantified by calculating computing ratio of nodes. Second algorithm overcomes the dynamic data load balancing problem. The default Hadoop job scheduler FIFO degrades the performance on heterogeneous cluster. A scheduling algorithm LATE (Longest Approximate Time to End) is proposed by M. Zaharia et al. [24] that can improve the Hadoop response time by a factor of 2 in a cluster of 200 virtual machines on Amazon's Elastic Compute Cloud (EC2). This algorithm provides the solution to how to robustly perform speculative execution for maximizing performance. LATE scheduler does not focus on the problem resulting from the phenomenon of dynamic loading that is addressed by LA (Load-Aware) scheduler proposed by Hsin-Han You et al. [25]. Faraz Ahmad et al. [26] proposed 'Tarazu', a suite of optimizations to improve MapReduce performance in heterogeneous clusters. Tarazu consists of a set of three schemes which are Communication-Aware Load Balancing of Map computation (CALB) across the nodes, Communication-Aware Scheduling of Map computation (CAS) to avoid network traffic, and Predictive Load Balancing of Reduce computation (PLB) across the nodes.

A white paper on Hadoop performance tuning [27] explains the tuning of various configuration parameters of Hadoop cluster, which directly affects the performance of a MapReduce job. Some major parameters described in that paper are block size, mapper's output compression, speculative execution, maximum map/reduce tasks, buffer size for sorting, temporary space and JVM tuning.

## IV. EXPERIMENTAL PLATFORM

A local Apache Hadoop-2.6.0 heterogeneous cluster consisting of five nodes is installed and configured. One node is fully devoted to NameNode (NN) and remaining four nodes serve as DataNodes (DNs). Cluster has both type of nodes physical and virtual as well as with different number of cores and RAM but all are running Ubuntu 14.04. Five nodes cluster has been installed using four physical machines. VMware is used to create virtual machine environment. Table I shows the architecture of physical machines used in cluster while Table II shows the configuration of the heterogeneous cluster with description of each node in cluster.

TABLE I. ARCHITECTURE OF MACHINES USED IN CLUSTER

| Machine | CPU Type | # Cores | RAM | Operating System |
|---|---|---|---|---|
| Machine A | Intel Xenon E5-2620 @ 2.10 GHz | 2 × 6 = 12 | 16 GB | Window 7 |
| Machine B | Intel Xenon E5504 @ 2.00 GHz | 1 × 4 = 4 | 2 GB | Ubuntu 14.04 |
| Machine C | Intel Xenon E5504 @ 2.00 GHz | 1 × 4 = 4 | 2 GB | Ubuntu 14.04 |
| Machine D | Intel Xenon E5-2630 @ 2.30 GHz | 2 × 6 = 12 | 32 GB | Window Server |

TABLE II. CONFIGURATION OF HETEROGENEOUS HADOOP CLUSTER

| Node | Node Type | Hosted on | # Cores | RAM |
|---|---|---|---|---|
| NameNode (NN) | Virtual | Machine A | 4 | 4 GB |
| DataNode1 (DN1) | Physical | Machine B | 4 | 2 GB |
| DataNode2 (DN2) | Physical | Machine C | 4 | 2 GB |
| DataNode3 (DN3) | Virtual | Machine D | 4 | 4 GB |
| DataNode4 (DN4) | Virtual | Machine D | 4 | 4 GB |

All the algorithms are implemented using JAVA and MapReduce 2.0 APIs. Hadoop-2.x version has introduced an improved and optimized framework MapReduce 2.0 (MRv2). MRv2 also known as *NextGen MapReduce or YARN (Yet Another Resource Negotiator)* [28] which controls job scheduling and manages cluster resources. Experiments were carried out on two click-steam datasets BMS_WebView_1 and BMS_WebView_2 from a web store [29]. Default setting of number of Mappers and Reducers are 12 and 4 respectively; it is explicitly stated whenever it is changed.

## V. FACTORS AFFECTING THE PERFORMANCE

As we have mentioned in earlier section that the performance of a MapReduce job can be enhanced either by improving the algorithm running as job or by fine tuning of various parameters specific to cluster. An algorithm showing good performance on homogeneous cluster drastically becomes poor on heterogeneous cluster. In this section we have discussed various factors that influence the performance of MapReduce based Apriori on homogeneous and heterogeneous Hadoop cluster. We have also applied techniques against these factors to improve the performance.

### A. Data Structures and Filtered Transactions

These two techniques are incorporated in the Apriori algorithm. Hash tree and trie (prefix tree) [30] are the central data structures in Apriori algorithm. F. Bodon [31] proposed the hash table trie data structure by applying hashing techniques on trie. Hash table trie was promising theoretically but failed to perform experimentally. Trie is the best performing data structure for the sequential implementation of Apriori; significant influence of the data structures can be found in [30-31]. In our earlier study [32], we have investigated the influence of the three data structures on MapReduce based Apriori when executed on our local Hadoop cluster. Experimental results showed that hash table trie drastically outperforms trie and hash tree with respect to execution time. Here we will represent only a part of its experimental results and compare with the filtered transactions method.

Transaction filtering was first used by C. Borgelt [33] in efficient sequential implementation of Apriori. If *t* is a transaction of database then a filtered transaction of *t* is the itemset obtained by removing infrequent items from transaction *t*. Filtered transactions are sufficient to determine all the frequent itemsets [34-35]. Here we only mention the pseudo codes for Apriori with filtered transactions. Algorithms from 1 to 7 depict the pseudo code of the driver class that contains three jobs and pseudo codes of Mapper, Reducer and Combiner classes of the three jobs. Algorithm 8 depicts the pseudo code of filtered transaction method.

### Algorithm 1. DriverApriori

```
// Find frequent 1-itemset L₁
Job1: //submitted single time
    OneItemsetMapper
    ItemsetCombiner
    ItemsetReducer
end Job1
// Find filtered transactions
JobFT: //submitted single time
    FT-ItemsetMapper
    ItemsetCombiner
    FT-ItemsetReducer
end JobFT
// Find frequent k-itemset Lₖ
for (k = 2; L_{k-1} ≠ ϕ; k++)
    Job2: //submitted multiple times
        K-ItemsetMapper
        ItemsetCombiner
        ItemsetReducer
    end Job
end for
```

### Algorithm 2. OneItemsetMapper, k = 1

**Input:** a block $b_i$ of database
key: byte offset of the line,
value: a transaction $t_i$
for each $t_i \in b_i$ do
    for each item $i \in t_i$ do
        write (i, 1);
    end for
end for

### Algorithm 3. ItemsetCombiner

key: itemset,
value: key's value list
for each key k do
    for each value v of k's value list
        sum += v;
    end for
    write(k, sum)
end for

### Algorithm 4. ItemsetReducer

key: itemset,
value: key's value list
for each key k do
    for each value v of k's value list
        sum += v;
    end for
    if sum >= min_supp_count
        write(k, sum)
    end if
end for

### Algorithm 5. FT-ItemsetMapper

**Input:** a block $b_i$ of database and $L_{k-1}$
key: byte offset of the line,
value: a transaction $t_i$
read frequent items from cache file in $L_1$
// $L_1$ may be a trie or hash table trie
for each $t_i \in b_i$ do
    $F_t$ = filterTransaction($L_1$, $t_i$); // Ft is the filtered transaction
    write ($F_t$, 1);
end for

### Algorithm 6. FT-ItemsetReducer

key: itemset,
value: key's value list
for each key k do
    for each value v of k's value list
        sum += v;
    end for
    write(k, sum)
end for

### Algorithm 7. K-ItemsetMapper, k ≥ 2

**Input:** a block $b_i$ of set of filtered transactions and $L_{k-1}$
key: byte offset of the line,
value: a filtered transaction $t_i$
read frequent (k-1)-itemsets from cache file in $L_{k-1}$
// $L_{k-1}$ may be a trie or hash table trie
$C_k$ = apriori-gen($L_{k-1}$); // $C_k$ may be a trie or hash table trie
for each $t_i \in$ block $b_i$ do
    occurrences = removeAtEnd($t_i$);
    $C_t$ = subset($C_k$, $t_i$); // $C_t$ may be a List
    for each candidate $c \in C_t$ do
        write (c, occurrences);
    end for
end for

### Algorithm 8. Filtered Transaction

filterTransaction($L_1$, $t_i$)
// parameters: trie $L_1$ with frequent items and a transaction $t_i$
// return value: filteredItems
    filteredItem = ""; // empty string
    for each item i of transaction $t_i$
        for each child node c of root of $L_1$
            if(i < c)
                break;
            else if(i > c)
                continue;
            else // append item to filteredItems
                filteredItems = filteredItems + " " + i;
                break;
        end for
    end for
    return filteredItems;
end filterTransaction

We have introduced a job for transaction filtering named as JobFT in between the Job1 and Job2. Job1 generates frequent 1-itemsets $L_1$ and Job2 generates frequent k-itemsets $L_k$ in $k^{th}$ iteration for k ≥ 2. FT-ItemsetMapper of JobFT reads $L_1$ from distributed cache and input file from HDFS. For each transaction $t_i$ it checks against $L_1$ to filter infrequent items from $t_i$ and produces $(F_t, 1)$ as key-value pairs where $F_t$ is the filtered transaction. FT-ItemsetReducer sums up the values associated with same filtered transaction and produces filtered transaction with its occurrence frequency. JobFT produces transactions with its occurring frequency in HDFS. K-ItemsetMapper of Job2 reads frequent itemsets $L_{k-1}$ from distributed cache and filtered transactions from HDFS. The method *removeAtEnd()* modifies the transaction by removing the occurrence frequency as well as returns this occurrence frequency. Candidates with this occurrence frequency are produced as key-value pairs. We have examined the algorithms for both data structures i.e. trie and hash table trie. ItemsetCombiner is same for all the three jobs since it makes the local sum on one node. ItemsetReducer makes the sum of local counts of the candidates received from all the nodes, checks count against minimum support threshold and produces candidates and its count as key-value pairs.

We have executed the MapReduce based Apriori using data structure trie and hash table trie (HTtrie) without filleted transactions and then with filtered transactions. Figures 1 and 2 show the execution times corresponding to trie, hash table trie (HTtrie), trie on filtered transactions (TrieOnFT) and hash table trie on filtered transactions (HTtrieOnFT) for datasets BMS_WebView_1 and BMS_WebView_2 respectively.

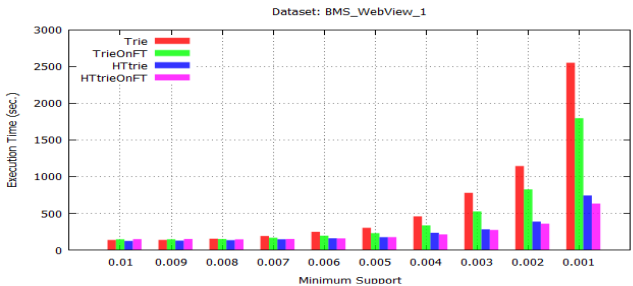

Fig. 1. Execution time of four variations of Apriori on BMS_WebView_1

From both the Fig. 1 and Fig. 2 we can see that the two techniques hash table trie and filtered transactions drastically reduce the execution time when applied independently and adds more improvement when applied jointly.

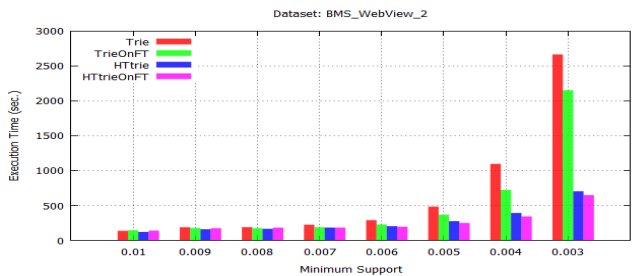

Fig. 2. Execution time of four variations of Apriori on BMS_WebView_2

### B. Speculative Execution

Speculative execution is a strategy of Hadoop that provides fault tolerance and reduces job execution time [10]. When a TaskTracker (a daemon process running on slave nodes that executes map and reduce tasks) performs poorly or crashes, the JobTracker (a demon process running on master node that accepts job and submit tasks to TaskTrackers) launches another backup task on another nodes to accomplish the task faster and successfully. This process of redundant execution is called *speculative execution* and the backup copy of task is called *speculative task* [25]. Among the original and speculative tasks which one completes first is kept while other is killed since it's no longer needed. Speculative execution is good at most of the time but it affects cluster efficiency by duplicating tasks. So it should be disabled (it is enabled by default) when a cluster has limited resources. Speculative execution is best suited in homogeneous environments but degrades the job performance in heterogeneous environments [24]. One can disable speculative execution for Mappers and Reducers by setting the value of "mapreduce.map.speculative" and "mapreduce.reduce.speculative" to *false* either in *mapred-site.xml* file of cluster configuration or in job configuration of MapReduce code [10].

We have executed the MapReduce based Apriori with trie data structure when speculation is on and off. Fig. 3 shows the observed execution time on dataset BMS_WebView_1 for varying value of minimum support.

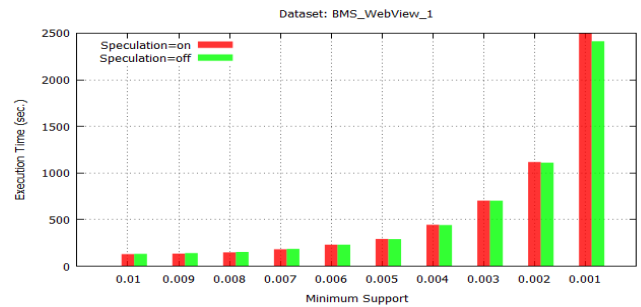

Fig. 3. Execution time of trie based Apriori when speculative execution is enabled and disabled

In Fig. 3 it can be seen that speculative execution comes in action only for a job taking longer time to complete. Speculative task is not launched for short jobs. It is launched when all the tasks of a job is assigned resources and then for a task running for a longer time or failed to complete.

### C. Detection and Elimination of Slower Nodes

Nodes in a heterogeneous cluster are of different capability. Heterogeneity may arise due to differences in hardware as well as using virtualization technology. Virtualization facilitates efficient utilization of resources and environments for different operating systems. There are many benefits of VM-hosted (virtual machine hosted) Hadoop such as lower cost of installation, on demand cluster setup, reuse of remaining physical infrastructure, on demand expansion and contraction of cluster size [36]. In our local cluster (Table 1 and 2) NN (NameNode), DN3 (DataNode3) and DN4 (DataNode4) are virtual machines while DN1 (DataNode1) and DN2 (DataNode2) are physical machines. However virtualization provides efficient re-use and management of resources but on the cost of performance. Virtual machines are slower in comparison to physical machines.

We have observed that eliminating slower node from cluster improves the performance. Detecting a slower node is not obvious always. We have used the idea of measuring heterogeneity in terms of data processing speed proposed in [23], where same MapReduce job with same amount of data is separately executed on each node and running time of each node is recorded. We have not executed the job separately on each node but on whole cluster. We have executed the job with 12 Mappers corresponding to frequent 2-itemsets generation for higher and lower value of minimum support on BMS_WebView_1 dataset. Fig. 4 and Fig. 5 show the snapshot of execution time of 12 Mappers on 4 DNs of the cluster for two values of minimum support.

![Fig 4 table]

Fig. 4. Snapshot of execution time of 12 Mappers on 4 DNs for higher value of minimum support

![Fig 5 table]

Fig. 5. Snapshot of execution time of 12 Mappers on 4 DNs for lower value of minimum support

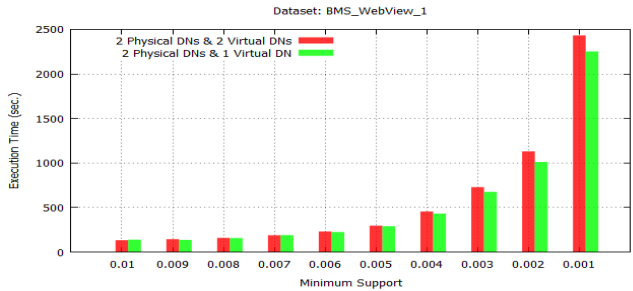

Fig. 6. Execution time of trie based Apriori on two different clusters

In Fig. 4 and 5, the DN on which the particular task is executed can be found by exploring the task's link. In Fig. 4, tasks showing elapsed time 28, 29 and 30 sec. are on VMs DataNodes, DN3 and DN4 whereas tasks showing elapsed time 19 and 20 sec. are on physical DataNodes, DN1 and DN2. Similar difference can be seen in Fig. 5 where tasks showing around 15 minutes are on VMs DataNodes and around 8 minutes are on Physical DataNodes. Removing both DN3 and DN4 greatly slowdown the performance but removing only one of them significantly enhances the performance. Fig. 6 shows the relative influence.

A natural question may arise here that is it a better idea to remove a slower DN? Here we would like to mention the two perspectives with Hadoop: developer's perspective and user's perspective. As a Hadoop developer one can design and incorporate an efficient data distribution scheme that assigns data blocks to DNs as per ratio of their processing speed. Similarly an efficient load balancing scheme can be incorporated to migrate tasks from busy or slower DNs to idle or faster DNs. Our case is user's perspective where we only focus of MapReduce based applications not on underlying algorithms of Hadoop system. So it's not a bad idea to eliminate slower DN if that reduces the execution time.

### D. Data Locality and Data Block Distribution

HDFS breaks a large file into smaller data blocks of 64 MB and stores as 3 replicated copies on DNs of cluster. Data blocks are the physical division of data. Data file can be logically divided using input split. The number of Mappers to be run for a job is equal to the number of logical splits. If input split is not defined then the default block size is the split size. In all the earlier cases discussed above, we have specified the split size which resulted into 12 Mappers and input file was not divided physically. In this case we have not used split size instead divide the input file into smaller blocks of 200 KB. So for the dataset BMS_WebView_1 there are 5 data blocks block0, block1, block2, block3, block4. This requires 5 Mappers to being executed corresponding to these 5 data blocks. When an input file is put into HDFS, it is automatically splitted into blocks and distributed to different DNs. Which block will allocate to which DN is not in user's control. Each time when putting the input file into HDFS, results a different distribution. We set the replication factor (RF) to 1 so that a block resides on only one DN. We put the same input file into HDFS twice and get two different block distributions (BD) BD1 and BD2. Now we set the replication factor to 4 so that all blocks would be available on each DN. Table III describes the three block distributions BD1, BD2 and BD3. The influence of three block distributions on the execution time of trie based Apriori for different value of minimum support is shown in Fig. 7.

TABLE III. THREE BLOCK DISTRIBUTIONS OF SAME FILE ON 4 DNs

| BD | RF | Blocks on DN1 | Blocks on DN2 | Blocks on DN3 | Blocks on DN4 |
|---|---|---|---|---|---|
| BD1 | 1 | block1, block2 | block0, block3 | No Block | block4 |
| BD2 | 1 | block0, block3 | block2 | block4 | block1 |
| BD3 | 4 | All 5 blocks | All 5 blocks | All 5 blocks | All 5 blocks |

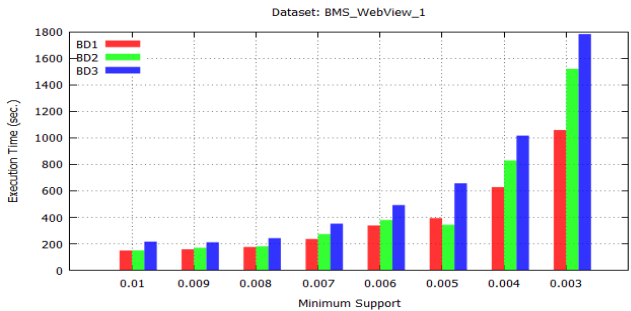

Fig. 7. Execution time of trie based Apriori on three different block distributions

In Fig. 7, BD1 exhibits minimum execution time compared to BD2 and BD3. In BD1, blocks are located only on three DNs i.e. two physical and one VMs DN. So in this case Mappers are not running on both slower DNs that make the execution faster. Execution time for BD2 is poor than that of BD1 due to using both VMs DNs. BD3 exhibits the worst performance since all the blocks are available on each DN. MapReduce processes a data block locally on the DN where the block is present. For block distribution BD3, all Mappers are running on the same DN since all blocks are locally available to each node. In different attempt of running a job, DN may be different each time but all the Mappers are being run on a same DN. All Mappers running on the same DN does not make use of available resources, which leads to increased execution time. Here it can be seen that due to higher replication factor data locality may be a hurdle that slow down the execution.

*E. Controlling Parallelism with Split Size*

Hadoop is designed to process big datasets that does not mean one cannot be benefited for small datasets. Apriori is a CPU-intensive algorithm and consumes a significant time for smaller datasets. To reduce the execution time we need more than one task running in parallel. Split is used to control the number of map tasks for a MapReduce job. A split may consist of multiple blocks and there may be multiple splits for a single block. So without changing the block size user can control the number of Mappers to be run for a particular job. We have used the method *setNumLinesPerSplit(Job job, int numLines)* of class *NLineInputFormat* from MapReduce library to set the number of lines per split. In our earlier cases we were running multiple Mappers against different parts of the same block. Here we set the split size 5K lines on block distribution BD3 which contains 5 data blocks. This creates 12 splits i.e. 12 Mappers over 5 blocks. If we don't define split size for BD3 then blocks are considered as input splits. Fig. 8 shows the difference in execution time for these two cases.

Here it can be seen that how the split size controls the parallelism. Smaller split size launches more number of Mappers which consequently increase the parallelism. It does not mean that more number of Mappers always results into better performance. Increasing the number of Mappers beyond a particular point starts to degrade the performance due to unnecessary overheads and shortage of resources [32]. To achieve the right level of parallelism it must be taken care that the map task is CPU-intensive or CPU-light as well as the size of dataset to be processed.

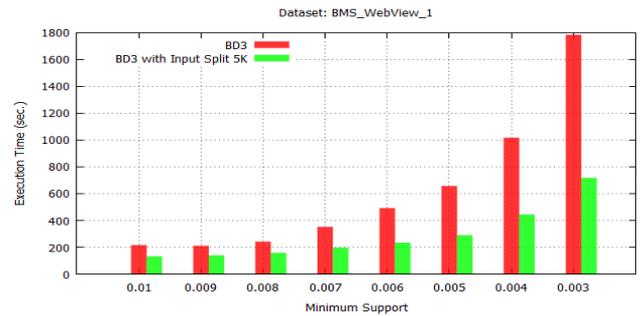

Fig. 8. Execution time of trie based Apriori with Input Split and without Input Split on BD3

*F. Issues Regarding MapReduce Implementation*

The efficiency of an algorithm running as a MapReduce job is extensively influenced by data structure used and algorithm itself. A third factor that cannot be ignored is the implementation technique. Implementation technique may be regarding to implementation of various modules of Apriori (e.g. candidate generation, support counting of candidates against each transaction, pruning of infrequent itemsets) or regarding to MapReduce implementation of Apriori. MapReduce implementation of Apriori is central to discussion here. A major issue in MapReduce based Apriori is to invoke candidate generation i.e. *apriori-gen()* (Algorithm 7) at appropriate place inside Mapper class. In our implementations we have invoked *apriori-gen()* inside customized method *map()* of Mapper class. In Mapper class, two methods *setup()* and *map()* are customized and one method *apriori-gen()* is defined. Method *setup()* is called once at the beginning of a task. It is customized to read frequent itemsets of previous iteration from distributed cache and to initialize prefix tree. Method *apriori-gen()* generates candidates using prefix tree containing frequent itemsets. The *map()* is invoked for each line of input split of dataset. If there are 100 lines of input assigned to a Mapper then *map()* method will be invoked for 100 times. Subsequently it invokes *apriori-gen()* repeatedly each time. Since *apriori-gen()* method produces candidates which is independent of input instance, so need not to invoke repeatedly inside *map()* method. The *apriori-gen()* method is computation intensive and increases the execution time when invoked repeatedly. This repeated computation can be fixed if we invoke *apriori-gen()* outside of *map()*. Theoretically it sounds good but did not work when invoked inside *setup()* method. We have also tried another way in which *apriori-gen()* is invoked inside overrided method *run()* of Mapper class but again could not achieve expected reduction in execution time.

## VI. CONCLUSIONS

In this paper, we have investigated a number of factors affecting the performance of MapReduce based Apriori algorithm on homogeneous and heterogeneous Hadoop cluster, and presented strategies to improve the performance. It has been shown that how hash table trie data structure and transaction filtering technique can significantly enhance the performance. Factors like speculative execution, physical & VMs DataNodes, data locality & block distribution, and split size are such that their proper tuning can directly enhance the performance of a MapReduce job even without making algorithmic optimization. Approaches of MapReduce implementation of Apriori is another important factor that also influence the performance.